\documentclass[aps,prl,floatfix,twocolumn,notitlepage,superscriptaddress,10pt]{revtex4-2}
\usepackage{xcolor}
\usepackage{grffile}
\usepackage{amsmath, amsthm, amssymb, bbold}
\usepackage[normalem]{ulem}
\usepackage{cancel}
\usepackage{bm}
\usepackage{microtype}
\usepackage{hyperref}
\usepackage{setspace}
\usepackage{grffile}
\hypersetup{colorlinks,linkcolor=blue,urlcolor=blue,citecolor=blue}
\usepackage{amsmath}    
\usepackage{amssymb}
\usepackage{graphicx}   
\usepackage{verbatim}   
\usepackage{color}      
\usepackage{hyperref}   
\usepackage[normalem]{ulem}
\usepackage{natbib}
\usepackage{enumitem} 
\usepackage{dsfont} 
\usepackage{lipsum}
\usepackage{mathtools}
\usepackage{soul}

\newcommand{\beq}{\begin{equation}}
\newcommand{\eeq}{\end{equation}}
\newcommand{\bea}{\begin{eqnarray}}
\newcommand{\eea}{\end{eqnarray}}

\newcommand{\be}{\begin{equation}}
\newcommand{\ee}{\end{equation}}

\definecolor{darkgreen}{rgb}{0,0.5,0}
\definecolor{orange}{rgb}{1,0.5,0}
\definecolor{grey}{rgb}{.6,.6,.6}

\newcommand{\tDelta}{\tilde \Delta}

\newcommand{\average}[1]{\langle #1\rangle}

\begin{document}

\title{Dynamic scaling and Family-Vicsek universality in $SU(N)$ quantum spin chains}
\author{C\u at\u alin Pa\c scu Moca}
\email{mocap@uoradea.ro}
\affiliation{Department of Physics, University of Oradea,  410087, Oradea, Romania}
\affiliation{Department of Theoretical Physics, Institute of Physics, Budapest University of Technology and Economics, M\H{u}egyetem rkp.~3, H-1111 Budapest, Hungary}
\author{Bal\'azs D\'ora}
\affiliation{Department of Theoretical Physics, Institute of Physics, Budapest University of Technology and Economics, M\H{u}egyetem rkp.~3, H-1111 Budapest, Hungary}
\author{Doru Sticlet}
\affiliation{National Institute for R\&D of Isotopic and Molecular Technologies, 67-103 Donat, 400293 Cluj-Napoca, Romania}
\author{Angelo Valli}
\affiliation{Department of Theoretical Physics, Institute of Physics, Budapest University of Technology and Economics, M\H{u}egyetem rkp.~3, H-1111 Budapest, Hungary}
\affiliation{HUN-REN—BME Quantum Dynamics and Correlations Research Group,
Budapest University of Technology and Economics, M\H{u}egyetem rkp. 3., H-1111 Budapest, Hungary}
\author{Toma\v{z} Prosen}
  \affiliation{Department of Physics, Faculty of Mathematics and Physics, University of Ljubljana, Jadranska 19, SI-1000 Ljubljana, Slovenia}	 
  \affiliation{Institute for Mathematics, Physics, and Mechanics, Jadranska 19, SI-1000 Ljubljana, Slovenia} 
\author{Gergely Zar\'and}
\affiliation{Department of Theoretical Physics, Institute of Physics, Budapest University of Technology and Economics, M\H{u}egyetem rkp.~3, H-1111 Budapest, Hungary}
\affiliation{HUN-REN—BME Quantum Dynamics and Correlations Research Group,
 Budapest University of Technology and Economics, M\H{u}egyetem rkp. 3., H-1111 Budapest, Hungary}

\date{\today}

\begin{abstract}
The Family-Vicsek scaling is a fundamental framework for understanding surface growth in 
non-equilibrium classical systems, providing a universal description of temporal surface roughness 
evolution. While universal scaling laws are well established in quantum 
systems, the applicability of Family-Vicsek  scaling in quantum many-body 
dynamics remains largely unexplored. Motivated by this, we investigate the 
infinite-temperature dynamics of one-dimensional $SU(N)$ spin chains, focusing on the 
well-known $SU(2)$ XXZ model and the $SU(3)$ Izergin-Korepin model. We compute the quantum analogue of classical surface roughness using the second cumulant of spin fluctuations and demonstrate universal scaling with respect to time and subsystem size. 
By systematically breaking global $SU(N)$ symmetry and integrability, we identify distinct transport regimes characterized by the dynamical exponent $z$: (i) ballistic transport with $z=1$, (ii) superdiffusive transport with the Kardar-Parisi-Zhang exponent $z=3/2$, and (iii) diffusive transport with the Edwards-Wilkinson exponent $z=2$. Notably, breaking integrability \emph{always} drives the system into the diffusive regime. Our results demonstrate that Family-Vicsek scaling extends beyond classical systems, holding universally across quantum many-body models with $SU(N)$ symmetry.
\end{abstract}
\maketitle

\paragraph{Introduction.}
The Family-Vicsek (FV) scaling~\cite{Vicsek1984,family1985scaling} describes universal dynamics that characterize the growth and roughness of surfaces in non-equilibrium systems~\cite{hohenberg1977theory}. Initially developed to model surface growth, the FV scaling law has since found applications in diverse areas, including fluid dynamics~\cite{dominguez2007scaling}, liquid crystals~\cite{Takeuchi2010}, biological growth~\cite{family1991dynamics, barabasi1995fractal}, and, more recently, quantum many-body systems~\cite{fujimoto2020family,fujimoto2021dynamical,glidden2021bidirectional,aditya2024family}. The core idea is to quantify surface or interface roughness as it evolves over time and system size. Surface roughness, typically represented by the width $W(l, t)$ of height fluctuations
of a segment $l$, exhibits power-law scaling
\begin{equation}\label{eq:width}  
    W(l, t) \sim l^\alpha f\left(\frac{t}{l^z}\right),  
\end{equation}  
where 
$t$ is time, $\alpha$  the roughness exponent, $\beta$  the growth exponent, and $z = \alpha/\beta$ is the dynamical exponent. The scaling function $f(x)$ governs the time evolution of roughness: in the early-time regime 
($t \ll l^z$, $f(x\ll1)\sim t^\beta$), $W(t) \sim t^\beta$ captures time-dependent roughening, while in the late-time regime ($t \gg l^z$, $f(x\gg 1)\sim {\rm const}$), roughness saturates as $W(t) \sim l^\alpha$, reflecting steady-state behavior~\cite{barabasi1995fractal}.  

Recently, there has been growing interest in exploring the universal scaling behavior established in classical surface growth models within the dynamics of quantum many-body systems~\cite{Nahum2017,glidden2021bidirectional,fujimoto2022impact,luis2022unveiling,jin2022kardar,bhakuni2024dynamic}. In this context, the height variable in classical models~\cite{barabasi1995fractal} corresponds to the expectation value of spin operators in quantum spin systems, while surface roughness relates to spin fluctuations. Although several models have been investigated, the primary focus was on non-interacting or weakly interacting systems~\cite{aditya2024family,fujimoto2020family}. 
Testing the FV scaling in strongly-correlated, interacting quantum systems remains largely unexplored.

SU($N$) spin models  constitute a fruitful playground for studying strongly correlated quantum systems, encompassing both integrable and nonintegrable regimes with rich transport phenomena. These models generalize conventional spin 
systems by incorporating higher-rank symmetries, making them valuable for exploring exotic quantum phases and 
non-equilibrium dynamics. A key example within this class is the $SU(2)$ spin-$\frac{1}{2}$ $XXZ$ model~\cite
{giamarchi2003quantum}, an interacting many-body system that serves as a cornerstone for understanding 
quantum magnetism and spin transport. Due to its integrability, the XXZ model exhibits distinct transport regimes, 
including ballistic, diffusive, and superdiffusive spin transport. The superdiffusive regime, characterized by 
Kardar-Parisi-Zhang (KPZ) scaling 
for two-point correlation functions~\cite{KPZ.86,Marko.2019,Wei.2022,Ye.2022,Rosenberg2024}, arises due 
to integrability and the presence of non-Abelian symmetries~\cite{Wei.2022,moca2023kardar}. Recent 
work has demonstrated that all integrable spin chains with non-Abelian $SU(N>2)$ symmetry~\cite{sutherland2004beautiful} exhibit superdiffusive transport with 
a dynamical exponent $z = 3/2$~\cite{ilievski2021superuniversality, Ye.2022}.
It is, however, puzzling that while correlations and diffusion are characterized by the KPZ exponents, the quantum spin models do not seem to belong to the KPZ universality class~\cite{Rosenberg2024,valli2024efficient}. 

So far, numerical investigations of FV scaling in these models remain challenging due to the difficulty of accessing large system sizes and 
timescales necessary for robust scaling analysis.  The recently introduced quantum generating function (QGF) approach~\cite{valli2024efficient} 
offers a powerful framework for addressing these challenges. By directly computing the generating function of conserved spin charges, the QGF method 
enables the efficient extraction of cumulants of the transferred spin $\Delta S^z_l$ in a segment $l$ as well as
their dynamical evolution~\cite{cecile2024squeezed} at or close to infinite temperature, and enables
the study of chain lengths of the order of hundreds or 
thousands of sites with high precision, across length- and time-scales that are inaccessible to other state-of-the-art methods.  
As we demonstrate using the QGF method for a number of spin models, $\Delta S^z_l$  obeys the scaling hypothesis in Eq.~\eqref{eq:width}.
%
with exponents $\alpha$ and $\beta$ satisfying the FV self-similar scaling relation $z = \alpha/\beta$ 
across all transport regimes. 

\paragraph{Key findings.} Within the XXZ model, we identify distinct transport regimes and construct the FV universal function for each case.  
In the easy-plane regime, where the anisotropy satisfies $\Delta < 1$, spin cumulants exhibit ballistic transport.
In this limit, the short-time behavior is consistent with FV scaling 
with exponents $\alpha=\beta = 1/2 $ and $z=1$. 
At the $SU(2)$ symmetric point, $\Delta = 1$, a KPZ scaling emerges with $z = 3/2$. In the easy-axis regime, $\Delta > 1$, the transport becomes diffusive-like~\footnote{Note that in integrable systems, $z=2$ does not imply normal diffusion as higher commulants typically exhibit critical scalings~\cite{KIP2022, KIP2024}} with an exponent $z = 2$.  
For the spin $S=1$ Izergin-Korepin (IK) model, we identify two distinct transport regimes. 
At the $SU(3)$ symmetric point, corresponding to $\tilde{\Delta} = 0$, KPZ scaling with $z = 3/2$ 
arises. However, breaking the $SU(3)$ global symmetry down to $U(1)$ leads to a transition to $z=2$.  
In both models, the introduction of integrability-breaking interactions~\cite{Surace.2023} universally 
induces a transition to a diffusive spin transport regime, characterized by an Edwards-Wilkinson (EW)
~\cite{Edwards1982} exponent $z = 2$, regardless of the anisotropy values. 
Moreover, the roughness exponent remains $\alpha = 1/2$ across all regimes of both models.

\paragraph{Quantum generating function.}
Recently, an efficient technique was introduced to calculate the QGF  
associated with a conserved $U(1)$ charge $\Sigma$~\cite{valli2024efficient}. 
The essence of the method is to calculate Heisenberg time evolution of the unitary operator 
$R_\Sigma(\lambda) = \exp(i\lambda\Sigma)$, defined in terms of the counting field $\lambda$, 
\begin{equation}\label{eq:time_evolution}
  R_\Sigma(\lambda, t) = U(t) R_\Sigma(\lambda) U^\dagger(t),
\end{equation}
where $ U(t)=\exp(-iHt)$ is the time-evolution operator. The generating function is then given by 
\begin{equation}\label{eq:G}
  G_\Sigma(\lambda,t) = \text{Tr} \average{R^{\phantom{\dagger}}_\Sigma(\lambda,t)R_\Sigma^\dagger(\lambda,0)}_{\tilde{\rho}},
\end{equation}
with $\average{...}_{\tilde{\rho}}$ representing the trace over the reduced density matrix $\tilde{\rho}$. Here we consider  infinite temperatures, corresponding to a density matrix proportional to unity.
The main numerical advantage is that for small values of $\lambda$, $ R(\lambda,t) $ remains close to the identity matrix and, 
therefore, it exhibits a slow growth of operator entanglement.
While in principle any cumulants can be extracted from the generating function, here we focus specifically on the second cumulant 
\begin{equation}
  \kappa_{2,\Sigma}(t) = -\partial_{\lambda}^2 G_\Sigma(\lambda, t)|_{\lambda\to 0}.
\end{equation}
In order to make contact with  surface growth models, we probe 
the magnetic fluctuations~\cite{fujimoto2021dynamical,cecile2024squeezed} within a finite interval $D_l$ of size $l$,
\begin{gather}
S^z_l = \sum_{j\in D_l} S^z_j, 
\end{gather}
and using the QGF approach, we investigate the dynamical scaling of $\kappa_{2,S^z}$ 
associated with the spin operator, $\Sigma=S_l^z$, 
that is the quantum equivalent of the classical height fluctuation introduced in Eq.~\eqref{eq:width}.

\begin{figure*}[t!]
	\centering
    \includegraphics[width=0.63\columnwidth]{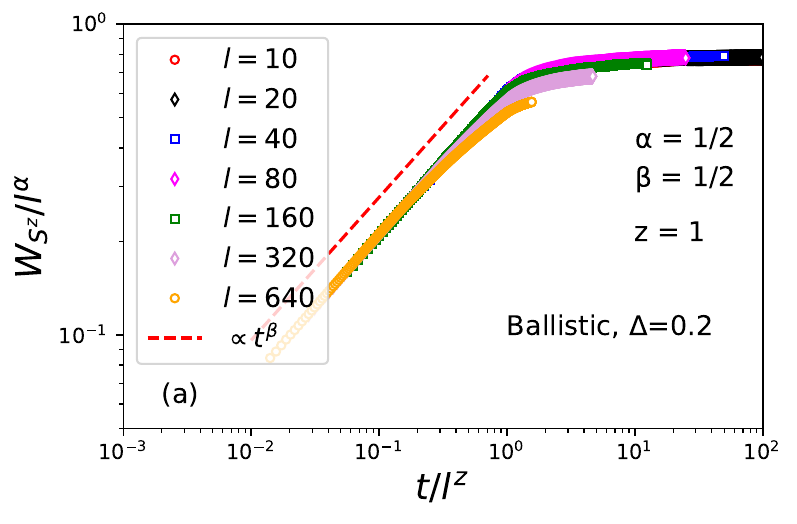}
    \includegraphics[width=0.6\columnwidth]{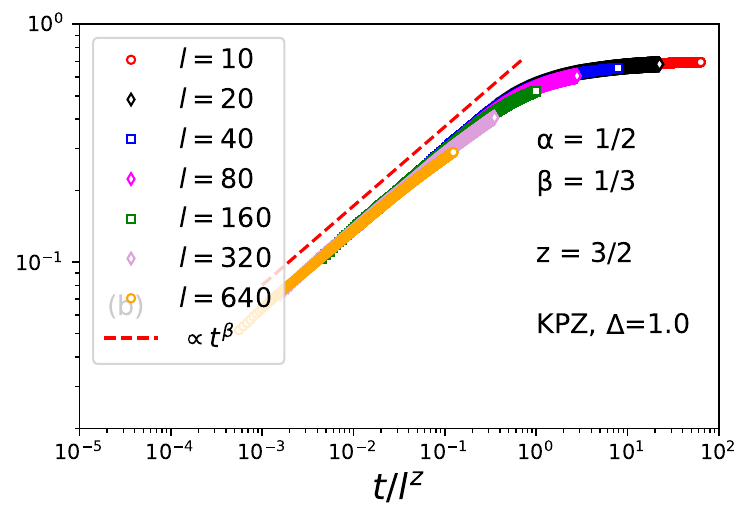}
    \includegraphics[width=0.6\columnwidth]{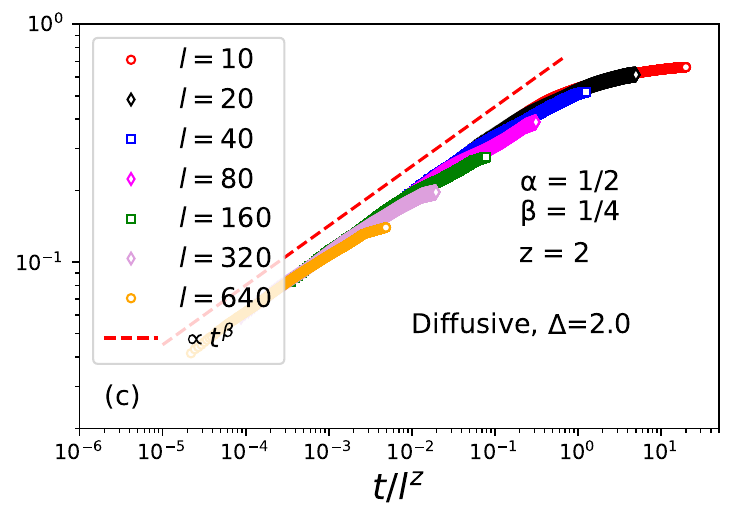}
   	\caption{The Family-Vicsek universal curves for the XXZ model are presented for $\Delta = \{0, 1, 2\}$, as indicated in each panel. For $\Delta < 1$ (left), the system exhibits ballistic scaling, while for $\Delta = 1$ (middle), it follows KPZ scaling, and for $\Delta > 1$ (right), it displays diffusive scaling. The system size is fixed at $L = 2000$ sites, with evolution carried out up to a maximum time of $t_{\rm max}\cdot J = 2000$. 
    In the ballistic limit, the deviations observed for larger system sizes are finite-size effects resulting from the restricted bond dimension employed. }
  	\label{fig:FV_SU2_int}
 \end{figure*}

\paragraph{Family-Vicsek universal scaling.}
In order to construct the FV universal functions, it is necessary to examine the scaling of the second cumulant  
with respect to the subsystem size. 
This is achieved by varying subsystem sizes $l$, located in the middle of the chain. 
Within the QGF approach, the initial spin operator is chosen as 
\begin{gather}
  R_{S^z}(\lambda) =\mathbb{1}\otimes\dots \mathbb{1}\otimes\underbrace{e^{i\lambda S^z_{-l/2}}\otimes \dots \otimes e^{i\lambda S^z_{l/2}}}_{-{l\over 2}<j\le {l\over 2}}\otimes\mathbb{1}\otimes\dots \mathbb{1}.
\end{gather}
%
The time evolution from  Eq.~\eqref{eq:time_evolution} is carried out using the time-evolving block decimation (TEBD) algorithm~\cite{Vidal.2003,Vidal.2004} implemented via the ITensor library~\cite{fishman2022itensor}. 
The generating function $G_{S^z}(\lambda, t)$ is computed using Eq.~\eqref{eq:G}. 
The second moments are obtained from the second-order Taylor expansion of $G_{S^z}(\lambda, t)$:
\begin{gather}
\kappa_{2,S^z_l}(t) \approx \frac{2}{\lambda^2} \left( 1 - G_{S^z_l}(\lambda,t) \right)+{\cal O}\left( \lambda^2 \right).
  \label{eq:moment}
\end{gather} 
We then compute the spatio-temporal dynamics of fluctuations as a function of subsystem size $l$ and time $t$ 
embedded in a system with L sites.
In the spirit of surface growth models, we define the roughness as 
\begin{gather}
W_{S^z}(l,t) = \sqrt{\kappa_{2,S^z_l}(t)},
\end{gather}
and construct the FV universal function $l^{-\alpha} W_{S^z}(l,t/l^{z})$.

\paragraph{XXZ model.} The Hamiltonian for the XXZ spin-$1/2$ chain with $L$ sites is given by  
\begin{gather}  
H_{\rm XXZ} = J \sum_{j=-L/2}^{L/2-1} \left( S_j^x S_{j+1}^x + S_j^y S_{j+1}^y + \Delta S_j^z S_{j+1}^z \right),  
\label{eq:H_XXZ}  
\end{gather}  
where $S_j^\alpha$ ($\alpha = x, y, z$) are spin-$1/2$ operators at site $j$, $J$ is the exchange coupling, and $\Delta$ is the anisotropy parameter. The first two terms describe spin-flip interactions in the $xy$ plane, while the third term represents the interaction along the $z$-axis, controlled by $\Delta$. Throughout this work, we set $J = 1$ as the energy unit.  

At the isotropic point $\Delta = 1$, the XXZ chain exhibits an $SU(2)$ symmetry, 
associated with the conservation of the total spin $\mathbf{S} = \sum_j \mathbf{S}_j$, 
with generators $S^+ = \sum_j S_j^+$, $S^- = \sum_j S_j^-$, and $S^z = \sum_j S_j^z$. 
This symmetry can be reduced to a $U(1)$ symmetry, corresponding to the conservation of 
the $S^z$ component of the total spin in the easy-plane ($\Delta<1$) or easy-axis ($\Delta>1$) regimes. 

The asymptotic values of the universal function can be understood intuitively through a straightforward argument. In the initial state under consideration, 
the infinite temperature ensures that the spins are completely uncorrelated, with equal probabilities $p_{\uparrow} = p_{\downarrow} = 1/2$ for both spin 
orientations. As the system evolves, the long-time limit leads to equal transition probabilities for a single 
spin, $p_{\uparrow\to \uparrow} = p_{\uparrow\to \downarrow} = p_{\downarrow\to \uparrow} = p_{\downarrow\to \downarrow} = 1/2$. 
This results in a second moment (variance) for a subsystem of size $l$  on the order of $\kappa_{2,S^z}(l,t\to \infty) = W_{S^z}^2(l) \approx l/2$, yielding the asymptotic 
scaling $W_{S^z}/l^\alpha \approx 1/\sqrt{2}$, in agreement with the numerical 
results~\footnote{A similar analysis for the IK model leads to the asymptotic value $W_{S^z}/l^\alpha \approx 2/\sqrt{3}$.}.
The XXZ model is integrable via the Bethe ansatz~\cite{babelon1983analysis,Karbach_1998}, 
but integrability can be broken by including additional interactions~\cite{Schagrin.2021,
Surace.2023}. To investigate the effects of integrability breaking, we introduce a 
next-nearest-neighbor interaction term,  
\begin{gather}  
H_{\rm nnn} = J_2 \sum_{j=-L/2}^{L/2-2} \mathbf{S}_j\cdot \mathbf{S}_{j+2}. 
\label{eq:H_nnn} 
\end{gather}  
The resulting Hamiltonian $H = H_{\rm XXZ} + H_{\rm nnn}$ describes the extended XXZ model.  

\begin{figure*}[t!]
  \centering
  \includegraphics[width=0.63\columnwidth]{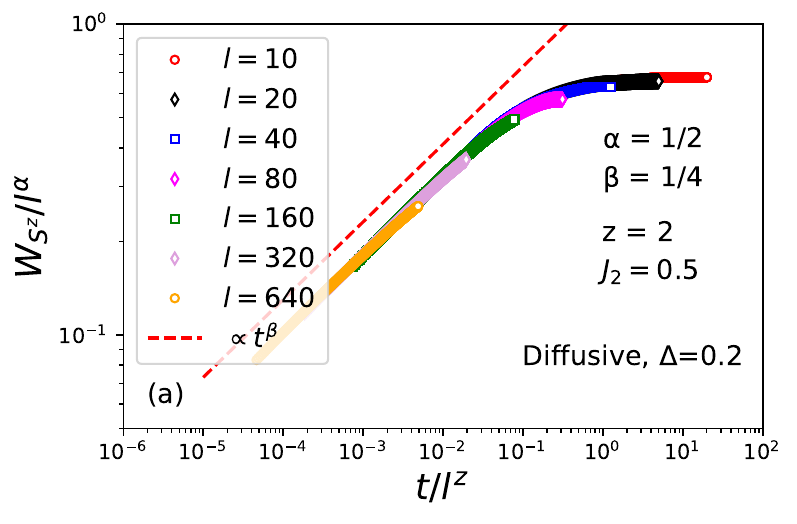}  
  \includegraphics[width=0.6\columnwidth]{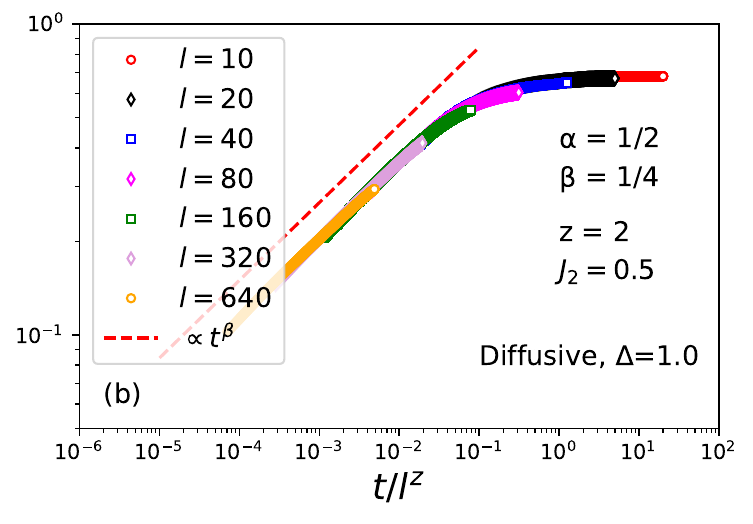}
  \includegraphics[width=0.6\columnwidth]{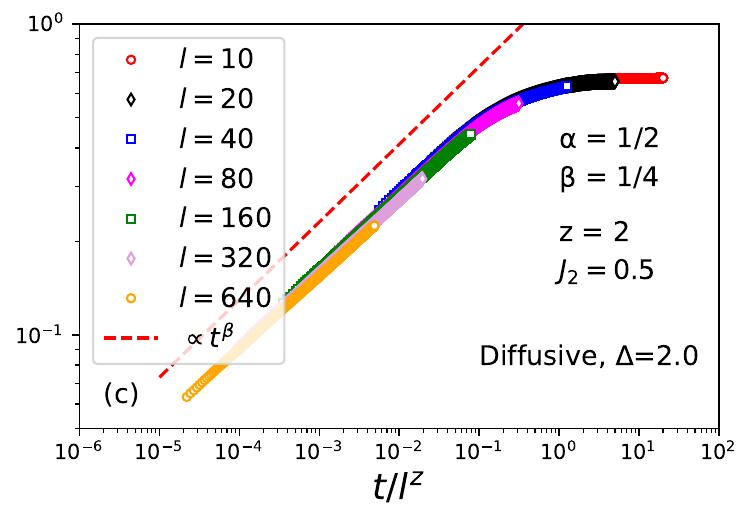}
   \caption{The Family-Vicsek universal curves for the non-integrable XXZ model with next-nearest-neighbor coupling $J_2$. Regardless of the value of $\Delta$, the system exhibits diffusive scaling. System size was fixed to $L=2000$ sites and $t_{\rm max}\cdot J=2000$.}
\label{fig:FV_SU2_non_int}
\end{figure*}

By rescaling the roughness with the exponent $\alpha=1/2$ and time with the dynamical exponent $z$, 
we demonstrate the universal collapse of the scaling function, and extract the corresponding exponent $\beta$ in each transport regime (see Figs.~\ref{fig:FV_SU2_int} and~\ref{fig:FV_SU2_non_int}).
Hence, the efficient evaluation of the cumulant  allows us to reach extremely large system sizes and an unprecedently accurate verification of the FV self-similar scaling.
In Fig.~\ref{fig:FV_SU2_int}(a) we consider the non-interacting case $\Delta=0$. 
In the absence of $\Delta$, the non-interacting spinons 
propagate ballistically $(z=1)$ and we find an early-time growth exponent $\beta = 1/2$~\cite{valli2024efficient}, whereas in the easy-axis regime, $\Delta>1$, the scaling is diffusive, corresponding to $(z=2)$ and $\beta = 1/4$ (Fig.~\ref{fig:FV_SU2_non_int}(c)). 
The latter exponent is easy to understand in terms of spin diffusion. At short times, change of spin occurs due to spin diffusion through the borders of the segment $l$, from a region $\Delta l \sim t^{1/2}$. Since the sign of the spins that diffuse through the boundary 
is random, this yields a change $|\Delta S^z_l|\sim t^{1/4}$.

For $\Delta = 1$, the model is integrable, exhibits a global $SU(2)$ symmetry, 
and is characterized by a superdiffusive (KPZ) scaling with dynamical exponent $z=3/2$.
In this regime, the KPZ  
prediction $\beta = 1/3$ is strongly supported by our numerical results in Fig.~\ref{fig:FV_SU2_int}(b). 
Breaking integrability, by including next nearest neighbor interaction terms as in Eq.~\eqref{eq:H_nnn}, 
brings the model in the diffusive regime, independently of whether $SU(2)$ symmetry is broken or not. 
The corresponding universal function is shown in Fig.~\ref{fig:FV_SU2_non_int}(a,b,c) for different values of $\Delta$. In this case, we always recover the diffusive FV growth exponents $\beta=1/4$, $\alpha=1/2$, and $z=2$. 

Notice that in all three regimes we have $\alpha=1/2$, which naturally follows from the infinite temperature environment: 
at very long times uncorrelated spins of random sign will populate the segment $l$ yielding $|\Delta S^z_l|\sim l^{1/2}$, 
that is $\alpha=1/2$.

\paragraph{Izergin-Korepin model.} 

\begin{figure}[t!]
  \centering
   \includegraphics[width=0.9\columnwidth]{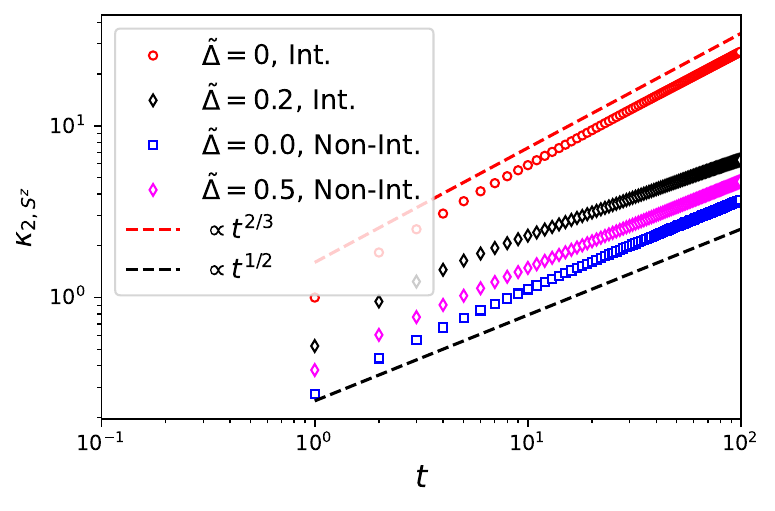}
   \caption{The early-time scaling of the second cumulant, $\kappa_{2,S^z}$, which characterizes magnetization fluctuations in half of the chain for the IK model, follows a superdiffusive KPZ ($z = 3/2$) scaling at $\tDelta = 0$, corresponding to the $SU(3)$ point. When symmetry is reduced or integrability is broken, it transitions to diffusive EW scaling ($z = 2$). In the integrable regime, a distinct crossover time emerges due to the small symmetry-breaking parameter $\tDelta = 0.2$. As the asymmetry increases, this crossover time shifts to smaller values. System size was fixed to $L=200$ sites.}
   \label{fig:early_time}
\end{figure}
\begin{figure*}[tb!]
	\centering
    \includegraphics[width=0.63\columnwidth]{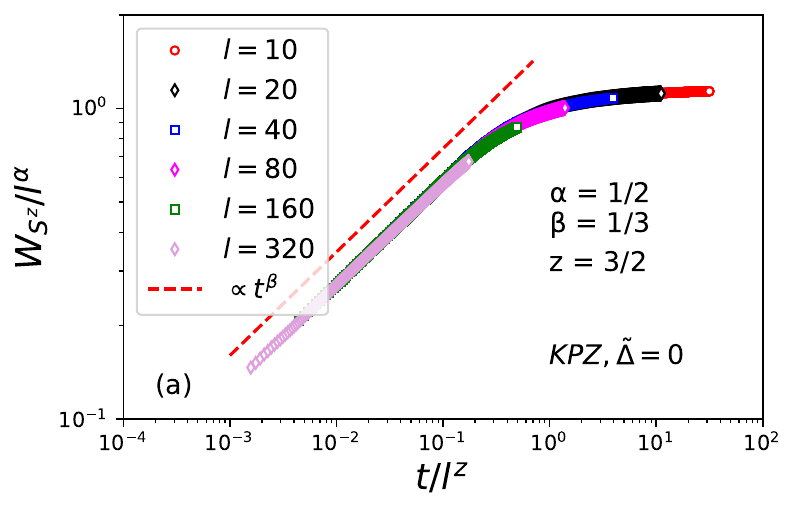}
    \includegraphics[width=0.61\columnwidth]{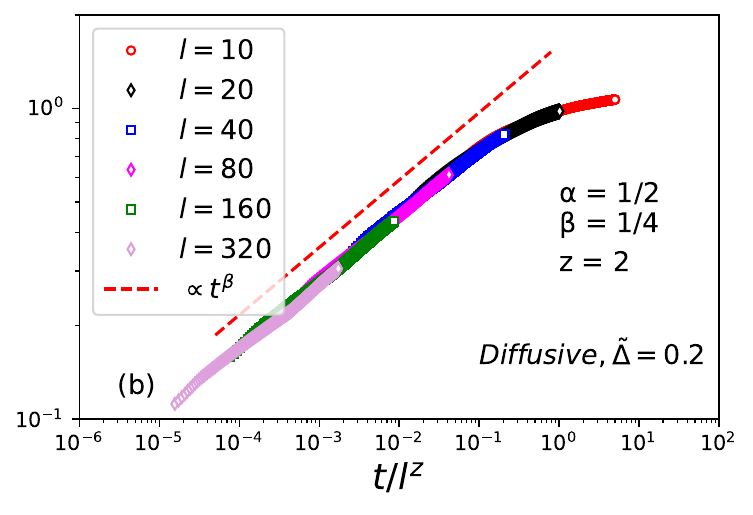}
    \includegraphics[width=0.58\columnwidth]{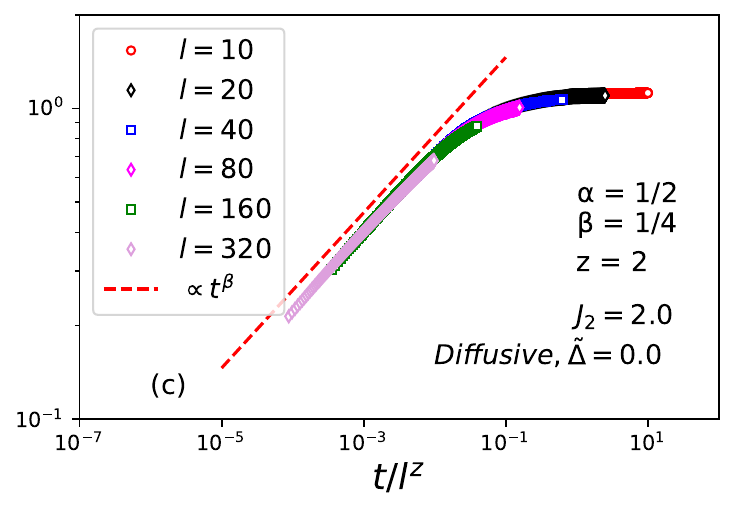}
   	\caption{(a, b) The Family-Vicsek universal curves for the integrable IK model at $\tDelta = 0$ and $\tDelta = 0.2$, as indicated in the panels. In the $\tDelta = 0$ case (a), the model exhibits a non-Abelian $SU(3)$ symmetry and follows KPZ scaling, whereas for $\tDelta \neq 0$, it transitions to diffusive scaling. (c) Breaking integrability at $\Delta = 0$ by introducing a next-nearest-neighbor interaction term ~\eqref{eq:H_nnn} drives the system into the diffusive regime of the Edwards-Wilkinson universality class. The system size is fixed at $L = 1000$ sites and $t_{\rm max}\cdot J =1000$.}
  	\label{fig:FV_SU3_int}
 \end{figure*}
 
 In the following, we investigate the anisotropic $SU(3)$ Izergin-Korepin (IK) model, with the Hamiltonian given in ~\ref{sec:IK}~\cite{izergin1981inverse, Ye.2022}. This model describes a system of spin-$S=1$ degrees of freedom, 
 where the local spin operators at site $i$ are defined by using the local operators $E_i^{\alpha,\beta} = |\alpha\rangle_i {}_{i}{\langle} \beta|$, with $\alpha, \beta \in \{1,2,3\}$. The anisotropy
 parameter $\tDelta$ governs both the interaction strength and the degree of symmetry breaking. Despite this tunability,
 the model remains integrable for any $\tDelta$ \cite{fan1997bethe} and possesses a global $U(1)$ symmetry associated 
 with the conservation of the $z$-component of the total spin, $S_z = \sum_{i} (E_i^{1,1}-E_i^{3,3})$.  

In the isotropic limit $\tDelta \to 0$, the model reduces to the bilinear-biquadratic Hamiltonian \cite{sutherland2004beautiful}, which exhibits a global non-Abelian $SU(3)$ symmetry corresponding to the conservation of the total spin-$S=1$. In terms of conventional spin operators, this limiting case takes the form  
\begin{equation}
H_{SU(3)} =J \sum_{i=-L/2}^{L/2-1} \left( \mathbf{S}_{i}  \cdot \mathbf{S}_{i+1} + (\mathbf{S}_i \cdot \mathbf{S}_{i+1})^2 \right).\label{eq:SU3}
\end{equation}  
Similar to the XXZ model, introducing a next-nearest-neighbor interaction as in Eq.~\eqref{eq:H_nnn} breaks the integrability of the IK Hamiltonian, leading to a transition from anomalous to diffusive transport.
 
\paragraph{Dynamical exponent $z$ in the IK model.}

It is easier to  accurately determine the dynamic exponent $z$, by examining the early-time dynamics. For that we alter the structure of the unitary operator 
$R_{S^z}(\lambda)$ as
\begin{gather}
  R_{S^z}(\lambda) = \underbrace{e^{i\lambda S^z_{-L/2}}\otimes \dots \otimes e^{i\lambda S^z_{0}}}_{-{L\over 2}<j\le 0}\otimes
  \overbrace{\mathbb{1}\otimes\dots \otimes \mathbb{1} }^{0<j<{L\over 2}}.
\end{gather}
The generating function $G_{S^z}(\lambda, t)$ is computed similarly, using Eq.~\eqref{eq:G} and the second moment is extracted according to Eq.~\eqref{eq:moment}. 
In order to correctly capture the early-time dynamics and minimize boundary effects, we stop the time evolution at $tJ \le L/2$. 
By considering $L=200$, we can reach times up to $t\le 10^2$. 

Figure~\ref{fig:early_time} presents the time evolution of the second cumulant, which quantifies spin fluctuations across the interface that divides the system into two halves. Across all regimes, we observe a power-law behavior, $\kappa_{2,S^z} \propto t^{1/z}$, and determine the corresponding dynamical exponent from the asymptotic value of $z(t) = (d/d\log t) \log \kappa_{2,S^z}(t)$. 
At $\tDelta = 0$, where the model exhibits $SU(3)$ symmetry, the system  is in a superdiffusive regime with a KPZ dynamical exponent of $z=3/2$~\cite{KPZ.86,moca2023kardar}. 
Breaking the $SU(3)$ symmetry or integrability drives the system into a diffusive regime characterized by the EW dynamical exponent $z = 2$~\cite{edwards1982surface,fujimoto2021dynamical} with a crossover time $t^*$ controlled by $\tDelta$ or $J_2$, respectively~\cite{Dupont.2020} 

\paragraph{FV scaling for the IK model.}
In Fig.~\ref{fig:FV_SU3_int}, we present the results of the FV scaling collapse for the integrable IK model, described by the Hamiltonian ~\eqref{eq:H_IK}. At the $SU(3)$ point, the system follows KPZ scaling with the characteristic dynamical exponent $z = 3/2$. 
When a finite $\tDelta \neq 0$ is introduced, the global $SU(3)$ symmetry is broken, resulting in diffusive scaling with dynamical exponent $z=2$ regardless of the sign of $\tDelta$. 
Similarly, breaking integrability by including the next-nearest-neighbor Hamiltonian term~\eqref{eq:H_nnn} also drives the system into the diffusive regime, independent of the values of $\Delta$. \footnote{For small $J_2 \ll J$, KPZ scaling persists over an extended time interval, followed by a crossover to the diffusive regime at a characteristic transition time $t^*$. As $J_2$ increases, this crossover occurs at progressively earlier times. See also Ref.~\cite{Dupont.2020}}

Our findings align closely with recent results obtained for spin ladder systems~\cite{wang2025}. Furthermore, our results are in agreement with predictions from generalized hydrodynamics (GHD) \cite{Bertini.2016, Castro.2016, Gopalakrishnan2023, doyon.2023}, which support a similar scaling crossover and provide a robust framework for describing transport phenomena in integrable systems. 
Additionally, our findings resonate with recent studies on integrable classical spin models \cite{McRoberts2022, McCarthy.2024, McRoberts.2024}, which reveal a slower onset of diffusion when constraints preserving non-Abelian symmetries are taken into account. 
This suggests that the underlying symmetry structure of the model plays a crucial role in determining the dynamics, further emphasizing the universality of these scaling behaviors across both quantum and classical settings.

\paragraph{Conclusions and Discussions.}\label{sec:conclusions}

In this work, we explore Family-Vicsek (FV) universality in one-dimensional XXZ and IK quantum spin models at infinite temperature. Using the quantum generating function approach, we show that FV scaling naturally emerges in these strongly correlated systems, linking classical surface growth and quantum transport. Our findings extend FV universality beyond the $SU(2)$ case, highlighting its broader relevance in quantum statistical mechanics.  

By analyzing spin fluctuations via the second cumulant in the XXZ model, we identify transport regimes governed by anisotropy: ballistic ($z = 1$) in the easy-plane regime, KPZ scaling ($z = 3/2$) at the $SU(2)$ point, and diffusive scaling ($z = 2$) in the easy-axis regime. These universal behaviors are confirmed through subsystem size dependence and time evolution, with exponents satisfying the FV relation, $z = \alpha / \beta$. Extending our study to the $SU(3)$ symmetric spin-$S=1$ Izergin-Korepin model, we observe KPZ scaling at the $SU(3)$ point and a transition to diffusion upon breaking $SU(3)$ symmetry to $U(1)$.  

Additionally, we examine integrability breaking via next-nearest-neighbor interactions, finding that it universally drives the system into a diffusive regime, regardless of anisotropies $\Delta$ or $\tilde{\Delta}$. This establishes the Edwards-Wilkinson class as the generic scaling behavior in non-integrable quantum systems. Our results highlight FV universality as a fundamental principle in quantum dynamics, with implications for symmetry, integrability, and scaling laws. 

\begin{acknowledgments}
This work received financial support from the Romanian CNCS/CCCDI–UEFISCDI, under projects
number PN-IV-P1-PCE-2023-0159 and PN-IV-P1-PCE-2023-0987. This research was also supported by the National Research, Development and Innovation Office - NKFIH within the Quantum Technology National Excellence Program (Project
No. 2017-1.2.1-NKP-2017-00001), 
 NKFIH research grants K138606, K142179 and SNN139581,
 by the BME-Nanotechnology FIKP grant (BME FIKP-NAT),
the QuantERA `QuSiED' grant No. 101017733, European Research Council (ERC)
through Advanced grant QUEST (Grant Agreement
No. 101096208), and Slovenian Research and Innovation agency (ARIS) through the Program P1-0402 and
Grants N1-0219, N1-0368.
\end{acknowledgments}

\bibliography{references}

\section{The Izergin-Korepin model}\label{sec:IK}

The Izergin-Korepin (IK) model is an integrable quantum spin chain with $SU(3)$ symmetry, first introduced in the 
context of inverse scattering theory~\cite{izergin1981inverse} and later explored in various quantum transport studies~\cite{Ye.2022}.

The Hamiltonian of the anisotropic IK model can be expressed in terms of operators $ E_i^{\alpha,\beta} = |\alpha\rangle_i {}_{i}{\langle} \beta|$, with $\alpha, \beta \in \{1,2,3\}$. The coupling structure is determined by the anisotropy parameter $\tilde{\Delta}$, which controls the relative strength of different hopping and interaction terms, 
\begin{widetext}
  \begin{gather}
  H_{\rm IK} = \frac{J}{\cosh 3\tDelta \cosh 2\tDelta} \sum_{i=-L/2}^{L/2-1} \Bigg[ \cosh 5\tDelta \left( E_i^{1,1} E_{i+1}^{1,1} + E_i^{3,3} E_{i+1}^{3,3} \right)
  + \sinh 2\tDelta (\sinh 3\tDelta - \cosh 3\tDelta) \left( E_i^{1,1} E_{i+1}^{2,2} + E_i^{2,2} E_{i+1}^{3,3} \right)\nonumber \\
  + \sinh 2\tDelta (\sinh 3\tDelta + \cosh 3\tDelta) \left( E_i^{2,2} E_{i+1}^{1,1} + E_i^{3,3} E_{i+1}^{2,2} \right)
  + 2 \sinh \tDelta \sinh 2\tDelta \left( e^{-2\tDelta} E_i^{1,1} E_{i+1}^{3,3} + e^{2\tDelta} E_i^{3,3} E_{i+1}^{1,1} \right)\nonumber \\
  + \cosh \tDelta \left( E_i^{1,3} E_{i+1}^{3,1} + E_i^{3,1} E_{i+1}^{1,3} \right)
  + \cosh 3\tDelta \left( E_i^{1,2} E_{i+1}^{2,1} + E_i^{2,1} E_{i+1}^{1,2} + E_i^{2,2} E_{i+1}^{2,2} + E_i^{2,3} E_{i+1}^{3,2} + E_i^{3,2} E_{i+1}^{2,3} \right)\nonumber\\
  - e^{-2\tDelta} \sinh 2\tDelta \left( E_i^{1,2} E_{i+1}^{3,2} + E_i^{2,1} E_{i+1}^{2,3} \right) 
  + e^{2\tDelta} \sinh 2\tDelta \left( E_i^{2,3} E_{i+1}^{2,1} + E_i^{3,2} E_{i+1}^{1,2} \right)
  \Bigg].
  \label{eq:H_IK}
\end{gather}
\end{widetext}
The model remains integrable for any value of $\tilde{\Delta}$, and at the $SU(3)$ symmetric point ($\tDelta=0$), it 
reduces to the bilinear-biquadratic Hamiltonian introduced in Eq.~\eqref{eq:SU3}.  
As we demonstrate, at the $SU(3)$ symmetric point, the model exhibits superdiffusive transport characterized by 
Kardar-Parisi-Zhang (KPZ) scaling ($z = 3/2$), mirroring the behavior observed in the $SU(2)$-symmetric XXZ chain in 
the presence of non-Abelian symmetries. Tuning $\tilde{\Delta} \neq 0$ breaks the $SU(3)$ symmetry down to its $U(1)$
 and induces a transition from KPZ scaling to diffusive transport ($z = 2$). This suggests that FV universality 
may extend beyond the $SU(2)$ case to a wider class of integrable quantum models.  

Although the model is integrable in its unperturbed form, introducing next-nearest-neighbor interactions or external fields breaks integrability, leading to a generic diffusive transport regime consistent with the Edwards-Wilkinson class. This trend is consistent with findings in other integrable spin chains, reinforcing the idea that integrability-breaking perturbations universally suppress anomalous transport.

\end{document}